\documentclass[twocolumn,footinbib,preprintnumbers,floatfix,aps,prl,10pt]{revtex4-2}
\usepackage{amsmath,amssymb,graphicx,booktabs,bm,psfrag,color,slashed,mathtools}
\usepackage{hyperref}
\usepackage{blindtext}
\usepackage[dvipsnames]{xcolor}

\newcommand{\spac}{{\hspace{0.3mm}}}

\makeatletter 
    
\renewcommand\onecolumngrid{
\do@columngrid{one}{\@ne}%
\def\set@footnotewidth{\onecolumngrid}
\def\footnoterule{\kern-6pt\hrule width 1.5in\kern6pt}%
}

\renewcommand\twocolumngrid{
        \def\footnoterule{
        \dimen@\skip\footins\divide\dimen@\thr@@
        \kern-\dimen@\hrule width.5in\kern\dimen@}
        \do@columngrid{mlt}{\tw@}
}%

\makeatother    

\begin{document}
\preprint{MITP-24-066}
\preprint{August 19, 2024}

\title{Factorization restoration through Glauber gluons}

\author{Thomas Becher$^a$}
\author{Patrick Hager$^b$}
\author{Sebastian Jaskiewicz$^a$}
\author{Matthias Neubert$^{b,c}$}
\author{Dominik Schwienbacher$^a$}
\affiliation{${}^a$Institut f\"ur Theoretische Physik {\em \&} AEC, Universit\"at Bern, Sidlerstrasse 5, CH-3012 Bern, Switzerland\\
${}^b$PRISMA$^+$ Cluster of Excellence {\em \&} MITP, Johannes Gutenberg University, 55099 Mainz, Germany\\
${}^c$Department of Physics, LEPP, Cornell University, Ithaca, NY 14853, U.S.A.}

\begin{abstract}
\vspace{-3mm} 
We analyze the low-energy dynamics of gap-between-jets cross sections at hadron colliders, for which phase factors in the hard amplitudes spoil collinear cancellations and lead to double (``super-leading'') logarithmic behavior. Based on a method-of-regions analysis, we identify three-loop contributions from perturbative active-active Glauber-gluon exchanges with the right structure to render the cross section consistent with PDF factorization below the gap veto scale. The Glauber contributions we identify are unambiguously defined without regulators beyond dimensional regularization.
\end{abstract}

\maketitle

Factorization, the separation of physics effects associated with different scales, is a fundamental property of quantum field theory. Indeed, the basis for all perturbative calculations of scattering processes at hadron colliders is the factorization of cross sections into non-perturbative (long-distance) parton distribution functions (PDFs) and high-energy (short-distance) partonic cross sections computed in perturbation theory. Crucially, PDF factorization also entails the absence of low-energy interactions between the colliding hadrons in the high-energy limit.
A formal proof for PDF factorization has only been presented for the inclusive Drell-Yan cross section~\cite{Collins:1985ue}. Over the years, a number of authors have expressed doubts that it will be valid in general~\cite{Collins:2007nk,Gaunt:2014ska,Zeng:2015iba}. 
Indeed, the observed breakdown of collinear factorization for space-like collinear splittings~\cite{Catani:2011st,Forshaw:2012bi,Schwartz:2017nmr,Cieri:2024ytf,Henn:2024qjq,Guan:2024hlf} is often taken as an indication that PDF factorization might be violated in higher orders of perturbation theory. Super-leading logarithms (SLLs)~\cite{Forshaw:2008cq} in exclusive jet cross sections have the same origin. They are double-logarithmic effects arising from complex phases in the hard-scattering amplitudes, which break color coherence and threaten the cancellation of collinear divergences in the cross section. Since PDF evolution is single-logarithmic, the presence of SLLs necessitates the existence of low-energy interactions between the incoming partons, and the key question is whether this is a perturbative effect. Both collinear factorization breaking and the SLLs are associated with Glauber dynamics, whose cancellation was crucial in the factorization proof for the Drell-Yan process. 

Collinear factorization breaking and SLLs were discovered long ago~\cite{Forshaw:2008cq,Catani:2011st,Forshaw:2012bi}, but an all-order understanding of these effects is lacking. For SLLs important progress was achieved recently, and the all-order structure of the leading effects is now known for arbitrary processes~\cite{Becher:2021zkk,Becher:2023mtx,Boer:2024hzh}. 
Based on an analysis in Soft-Collinear Effective Theory (SCET)~\cite{Bauer:2001yt,Bauer:2002nz,Beneke:2002ph,Beneke:2002ni}, the separation of scales in the cross section was accomplished and a systematic framework for investigating factorization was provided. SLLs arise first at four-loop order, and the preservation of PDF factorization would require a highly intricate interplay of high-energy and (perturbative) low-energy dynamics at this order, whose precise mechanism has remained elusive.

In this Letter, we address this outstanding challenge and identify, for the first time, a contribution of an active-active Glauber exchange (a gluon exchange between partons participating in the hard scattering) to a cross section. We show that in leading-logarithmic approximation the result has exactly the required form to turn the double-logarithmic back into single-logarithmic evolution. While our analysis does not amount to a proof of PDF factorization, it demonstrates that the breaking of collinear factorization does not necessarily translate into a breaking of PDF factorization. On the technical side, our analysis relies on a method-of-regions~\cite{Beneke:1997zp,Smirnov:2002pj} computation of box and pentagon diagrams. Obtaining a complete list of the regions contributing to a given kinematic configuration is not a straightforward task and, indeed, an area of active research~\cite{Jantzen:2012mw,Ananthanarayan:2020ptw,Gardi:2022khw,Beneke:2023wmt,Ma:2023hrt,Guan:2024hlf,Smirnov:2024pbj,Gardi:2024axt}. Interestingly, the Glauber region relevant to our case was not identified in earlier literature and is missed by the available computer codes used to search for regions. This contribution is the first example of a {\em genuine\/} Glauber effect in active-active parton scattering, i.e.\ it is well-defined without additional regulators and is not contained within other regions.

\begin{figure}
\centering
\includegraphics{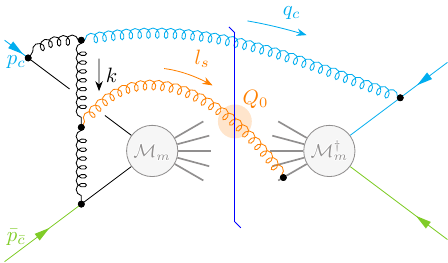}
\caption{Sample perturbative contribution to the gap-between-jets cross section. The gray inner subdiagrams make up the hard function $\bm{\mathcal{H}}_m$, while the remainder is part of $\bm{\mathcal{W}}_m$. The orange gluon is soft and enters the veto region, the blue and green partons are collinear to the beams. Possible scalings of the virtual gluon momentum $k$ will be analyzed below.}
\label{fig:diagram1}
\end{figure}

The observable we study is the production of $M$ jets in hadron-hadron collisions at large transverse momentum $Q$, together with a stringent veto on radiation emitted into a gap outside the jets, with an associated veto scale $Q_0$. Such gap-between-jets observables have been measured at the LHC~\cite{ATLAS:2011yyh} and are examples of non-global hadron-collider observables involving two disparate scales. For small $Q_0\ll Q$, one can derive a factorization theorem for such processes, which reads~\cite{Balsiger:2018ezi,Becher:2021zkk,Becher:2023mtx}
\begin{align}\label{hadronfact}
   &\sigma(Q_0) = \sum_{m=m_0}^\infty \int d \xi_1 d \xi_2 \\
   &\times \big\langle \bm{\mathcal{H}}_m(\{\underline{n}\},Q,\xi_1,\xi_2,\mu)
    \otimes \bm{\mathcal{W}}_m(\{\underline{n}\},Q_0,\xi_1,\xi_2,\mu) \big\rangle \,, \notag
\end{align}
where $m_0=2+M$ is the number of partons at Born-level, $\xi_i$ are the momentum fractions of the initial-state partons, and the sum includes all partonic subprocesses. The hard functions are the squared amplitudes for producing the energetic partons inside the jets, integrated over the energies of the final-state particles,
\begin{equation}\label{hardfun}
   \bm{\mathcal{H}}_m(\{\underline{n}\},Q,\xi_1,\xi_2,\mu) 
   = \int\!d\mathcal{E}_m\,|\mathcal{M}_m(\{\underline{p}\})\rangle 
    \langle\mathcal{M}_m(\{\underline{p}\})| \,,
\end{equation}
while keeping the parton directions $\{\underline{n}\}=\{n_1,\dots, n_m\}$ fixed. The explicit form of the energy integration can be found in (2.3) of~\cite{Becher:2023mtx}. The integration over the final-state parton directions is indicated by the symbol $\otimes$ in~\eqref{hadronfact}. The color indices of the hard partons are kept open and $\langle\dots\rangle$ denotes the color trace, which is taken after combining the hard functions with the low-energy matrix elements $\bm{\mathcal{W}}_m$, which contain the dynamics associated with the perturbative scale $Q_0$, as depicted in Fig.~\ref{fig:diagram1}, as well as non-perturbative QCD effects. The main result of our Letter is that, at least up to three-loop order, the perturbative part of $\bm{\mathcal{W}}_m$ is consistent with PDF factorization. 

The SLL analysis in~\cite{Becher:2021zkk,Becher:2023mtx} was based on the renormalization-group evolution of the hard functions from the high scale $\mu_h=Q$ to a low scale $\mu_s\sim Q_0$. The leading logarithms were obtained by iterating the one-loop anomalous dimension~\footnote{Note that the normalization of $\bm{\Gamma}^c$ and $\bm{V}^G$ differs from the one used in~\cite{Becher:2021zkk,Becher:2023mtx} by a factor 4.} 
\begin{equation}\label{eq:anomalous_dimension_soft_part}
   \bm\Gamma^{H} 
   = \gamma_{\rm cusp}(\alpha_s) \Bigl( \bm{\Gamma}^c
    \ln\frac{\mu^2}{Q^2} +\bm{V}^G \Bigr)
    + \frac{\alpha_s}{4\pi}\,\overline{\bm{\Gamma}} 
    + \bm{\Gamma}^C \,,
\end{equation}
where $\gamma_{\rm cusp}=\alpha_s/\pi+\dots$ is the light-like cusp anomalous dimension. The soft piece consists of $\bm{\Gamma}^c$ and $\bm{V}^G$, which account for soft$+$collinear emissions from one of the two initial-state partons and complex phases arising from virtual gluon exchange between them, respectively. $\overline{\bm{\Gamma}}$ corresponds to gluon emission into the gap, and $\bm{\Gamma}^C$ denotes purely collinear contributions. The anomalous dimension is an operator in color space and a matrix in the space of parton multiplicities $m$. An application of $\bm\Gamma^{H}$ can either increase the number of partons, corresponding to a real emission, or leave them unchanged for virtual terms. The SLLs originate from $\bm{\Gamma}^c$. Using simple identities among the various terms in \eqref{eq:anomalous_dimension_soft_part}~\cite{Becher:2021zkk}, one finds that the relevant color traces are of the form
\begin{equation}\label{SLL4}
   C_{rn} = \big\langle \bm{\mathcal{H}}^{(0)}_{m_0}\spac 
    \left( \bm{\Gamma}^c \right)^r \bm{V}^G \left( \bm{\Gamma}^c \right)^{n-r} \bm{V}^G\,\overline{\bm{\Gamma}}\otimes \bm{1}\big\rangle \,.
\end{equation}
Performing the associated scale integrals for evolution from $Q$ down to the scale $\mu_s\sim Q_0$ produces single logarithms for $\bm{V}^G$ and $\overline{\bm{\Gamma}}$, but double logarithms for $\bm{\Gamma}^c$. The color traces $C_{rn}$ thus contribute at order $\alpha_s^{n+3} L_s^{2n+3}$ in perturbation theory, where $L_s=\ln(Q/\mu_s)$. SLLs first arise at four-loop order and involve $C_{01}$ and $C_{11}$. In~\eqref{SLL4}, $\bm{\mathcal{H}}^{(0)}_{m_0}$ are the Born-level hard functions and we use that $\bm{\mathcal{W}}_m^{(0)}=\bm{1}$ at lowest order. 

The fact that the cross section $\sigma(Q_0)$ must be independent of the renormalization scale $\mu_s$ imposes non-trivial conditions on the low-energy matrix elements $\bm{\mathcal{W}}_m(\mu_s)=\bm{Z}\spac\bm{\mathcal{W}}_m^{\rm bare}$. The renormalization factor $\bm{Z}$ is related to the anomalous dimension~\eqref{eq:anomalous_dimension_soft_part}~\cite{Becher:2009qa}, and using its three-loop expression one finds that the leading UV poles in $d=4-2\varepsilon$ dimensions must be of the form
\begin{align}\label{eq:Wm_poles}
   \bm{\mathcal{W}}_m^{\rm bare}
   &= \bm{1} + \frac{\alpha_s}{4\pi}\,\frac{\overline{\bm{\Gamma}}}{2\varepsilon}
    + \left( \frac{\alpha_s}{4\pi} \right)^2 \left( \frac{\bm{V}^G\,\overline{\bm{\Gamma}}}{2\varepsilon^2} + \dots \right) \notag\\
   &\quad + \left( \frac{\alpha_s}{4\pi} \right)^3 \left( \frac{\bm{V}^G\spac\bm{V}^G\,\overline{\bm{\Gamma}}}{3\varepsilon^3}
    - \frac{\bm{\Gamma}^c\spac\bm{V}^G\,\overline{\bm{\Gamma}}}{3\varepsilon^3} \ln\frac{Q^2}{\mu_s^2} + \dots \right) \notag\\[1mm]
   &\quad + \mathcal{O}(\alpha_s^4) \,.
\end{align}
We only show terms which, after combining with the hard functions in~\eqref{hadronfact} and taking the color trace, produce contributions compatible with SLLs at four-loop order and beyond. Under the color trace, we can replace~\cite{Becher:2023mtx} 
\begin{align}   
    \bm{V}^G\,\overline{\bm{\Gamma}} 
    &\to 16i\pi\bm{X}_1 \,, \quad
    \bm{\Gamma}^c\spac\bm{V}^G\,\overline{\bm{\Gamma}} 
     \to 16i\pi\spac N_c\spac\bm{X}_1 \,, \notag\\
    \bm{V}^G\spac\bm{V}^G\,\overline{\bm{\Gamma}} 
    &\to - 6\pi^2 N_c\spac\bm{X}_2 \,,
\end{align}
where
\begin{equation}
   \bm{X}_1 = i f^{abc} \sum_{j>2} J_j\,\bm{T}_1^a\spac\bm{T}_2^b\spac\bm{T}_j^c \,, \hspace{2.5mm} \bm{X}_2 =\frac{1}{N_c}\sum_{j>2} J_j\,\bm{O}_1^{(j)} ,
\end{equation}
with $\bm{O}_1^{(j)}$ defined in (6.36) of~\cite{Becher:2023mtx}. The sums extend over all final-state partons $j>2$ in the Born-level process, and the angular integral $J_j$ has been given in (16) of~\cite{Becher:2021zkk}.

We now compute the perturbative part of $\bm{\mathcal{W}}_m^{\rm bare}$ order by order in $\alpha_s$ and check whether it matches the structure~\eqref{eq:Wm_poles}. The one-loop term $\propto\overline{\bm{\Gamma}}$ is the divergence associated with a soft exchange between hard legs and is obtained from soft Wilson-line matrix elements in the low-energy theory or, equivalently, by taking the product of two tree-level soft currents $\bm{J}^{a(0)}_\mu(l_s)$ and integrating the momentum $l_s$ over the gap region under the restriction $l_s^0<Q_0$. The $\alpha_s^2$ term $\propto\bm{V}^G\,\overline{\bm{\Gamma}}$ arises from real-virtual corrections to the same matrix elements. The complex phase in $\bm{V}^G$ is directly related to the imaginary part of the one-loop soft current $\bm{J}^{a(1)}_\mu$~\cite{Catani:2000pi}. To isolate the structure $\bm{V}^G\spac\bm{V}^G\,\overline{\bm{\Gamma}}$, we have analyzed the product $\bm{J}^{\mu,a(1)} \bm{J}^{a(1)\dagger}_\mu$ as well the product of $\bm{J}^{a(2)}_\mu$ (including the tripole terms) 
\cite{Duhr:2013msa,Dixon:2019lnw} with a tree-level current, comparing the results for space-like and time-like kinematics. From these computations, we conclude that all terms in~\eqref{eq:Wm_poles} other than the structure $\bm{\Gamma}^c\spac\bm{V}^G\,\overline{\bm{\Gamma}}$ are correctly reproduced through soft physics alone. This final term involves a logarithm of the hard scale $Q$, but the purely soft matrix elements are independent of $Q$. 

A $Q$ dependence in the low-energy theory can arise i) if it contains low-energy modes with virtualities below $Q_0$ (ultra-soft modes in SCET$_{\rm I}$ and soft-collinear modes in SCET$_{\rm II}$ are examples of this), or ii) via a collinear anomaly~\cite{Becher:2010tm,Chiu:2011qc}, which breaks a classical rescaling invariance of SCET and leads to rapidity divergences. These divergences cancel among the different sectors but leave behind a logarithmic $Q$ dependence. However, in our case the purely collinear matrix elements are scaleless even with a rapidity regulator because they correspond to ``partonic PDFs''. So even in scenario ii) the theory necessarily involves low-energy interactions between the soft and the two collinear sectors. Such an interaction can be mediated by Glauber gluons. A final scenario iii) would be that the $\bm{\Gamma}^c\spac\bm{V}^G\,\overline{\bm{\Gamma}}$ term is not reproduced by perturbative physics, but by non-perturbative interactions between soft and collinear particles. This option is incompatible with PDF factorization and would require a generalization involving non-perturbative two-nucleon matrix elements.

To identify the relevant mechanism and clarify whether it is perturbative or non-perturbative, we have performed a method-of-regions analysis of the three-loop QCD diagrams contributing to $\bm{\mathcal{W}}_m$, specifically those that can produce the structure $\bm{\Gamma}^c\spac\bm{V}^G\,\overline{\bm{\Gamma}}$. These graphs feature a soft-gluon emission into the gap, a collinear emission, and a virtual gluon exchange with, as of yet, unspecified kinematic scaling of its loop-momentum $k$.  In dimensional regularization, diagrams of this type vanish due to scalelessness, unless the soft emission is directly radiated off a virtual gluon connecting the collinear and anti-collinear sectors, as depicted in Fig.~\ref{fig:diagram1}. Two other relevant diagrams are obtained by attaching the virtual gluon to the upper quark line either before or after the collinear gluon emission. To elucidate the structure of the low-energy theory, we need to identify all kinematic regions of these diagrams. 

\begin{figure}
    \centering
    \includegraphics[]{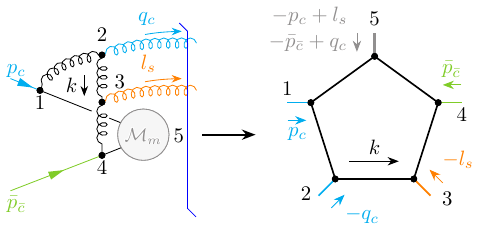}
   \includegraphics{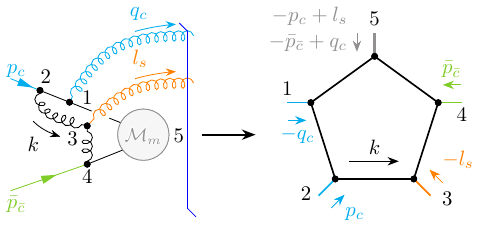}
    \caption{Mapping of low-energy contributions to $\bm{\mathcal{W}}_m$ onto pentagon diagrams. The external momentum $p_5^2\ne 0$ flows into the hard amplitude $\mathcal{M}_m$.}
    \label{fig:Pent2}
\end{figure}

We begin our investigation by stripping off the tensor structure of the numerators and considering the regions decomposition of the dimensionally regulated scalar integrals. As the scaling of the real emissions is restricted by the external kinematics, we focus on the loop integral over $k$, which can be mapped onto box and pentagon structures, as depicted in Fig.~\ref{fig:Pent2} for the latter case. This allows for a direct comparison of the regions results and the known full expressions, thus ascertaining that all regions are correctly identified. Introducing a small power-counting parameter $\lambda = Q_0/Q$, the external legs carry collinear momenta $p_c$, $q_c$ whose components scale as $(n\cdot p_c,\bar n\cdot p_c,p_{c\perp})\equiv(p_c^+,p_c^-,p_{c\perp})\sim Q(\lambda^2,1,\lambda)$, an anti-collinear momentum $\bar{p}_{\bar{c}}$ $\sim Q(1,\lambda^2,\lambda)$, and a soft momentum $l_s\sim Q(\lambda,\lambda,\lambda)$. We introduce two light-cone vectors $n$ and $\bar n$ (with $n^2=\bar n^2=0$ and $n\cdot\bar n=2$) along the directions of $p_c$ and $\bar{p}_{\bar{c}}$. In the following, we focus on the two pentagon structures, for which a complete set of invariants is given by $s_{i,i+1}=\left(p_i+p_{i+1}\right)^2$ and $m^2=p_5^2$ for inflowing external momenta $p_i$ associated with the external lines. At leading power in $\lambda$, they are given by (choosing $p_{c\perp}=\bar{p}_{\bar{c}\perp}=0$)
\begin{align}\label{eq:invariants_pentagon}
   s_{12} &= -p_{c}^- q_{c}^+ \,, & s_{23} &= q_{c}^- l_{s}^+ \,, & 
    s_{45} &= -(p_{c}^- -q_{c}^-)l_{s}^+ \,, \nonumber\\
   s_{34} &= -\bar{p}_{\bar{c}}^{\spac +} l_{s}^{-} \,,\hspace*{-0.15cm} & 
    s_{51} &= - q_{c}^{-} \bar{p}_{\bar{c}}^{\spac +} \,, \hspace*{-0.15cm}& 
   m^2 &= (p_{c}^{-} -q_{c}^{-}) \bar{p}_{\bar{c}}^{\spac +}
\end{align}
for the upper graph in Fig.~\ref{fig:Pent2}. For the lower graph $s_{23}=-p_c^- l_{s}^+$ and $s_{51}=p_c^- \bar{p}_{\bar{c}}^{\spac +}$, while all other invariants remain the same. Before studying the physical case, we consider Euclidean kinematics, where all $s_{i,i+1}<0$ and $m^2<0$. To identify the contributing regions, we utilize \texttt{pySecDec}~\cite{Heinrich:2021dbf} and translate the parameter-space output into momentum regions. At leading power in $\lambda$, the only non-zero contribution for both pentagon integrals stems from the soft-collinear region $k\sim Q(\lambda,\lambda^2,\lambda^{3/2})$~\cite{Becher:2003qh}. For example, the upper diagram in Fig.~\ref{fig:Pent2} corresponds to 
\begin{align}\label{eq:softcollinear_pentagon_euclidean}
   I^{\text{sc}} &= i(4\pi)^{2-\varepsilon}
    \int\frac{d^dk}{(2\pi)^d}\,\frac{1}{k^2+i0}\,
    \frac{1}{-l_{s}^+\spac k^{-}+i0}\,
    \frac{1}{q_c^-\spac k^++i0} \nonumber \\ 
   &\quad\times \frac{1}{\bar{p}_{\bar{c}}^{\spac +}(k^- -l_{s}^-)+i0}\,
    \frac{1}{\left[-(p_c^--q_c^-)\spac k^+-p_c^-q_c^+ +i0\right]} \nonumber\\[2mm]
   &= \frac{\Gamma^2(\varepsilon)\spac\Gamma(1-\varepsilon)}{s_{45}\spac s_{51}}\,\spac{}_2F_1(1,1;1-\varepsilon;1-\frac{m^2\spac s_{23}}{s_{45}\spac s_{51}}) \nonumber \\ 
   &\quad\times\left(-s_{12}\right)^{-1-\varepsilon} \left(-s_{34}\right)^{-1-\varepsilon} 
    \left(-m^2\right)^{1+\varepsilon} \,,
  \end{align} 
with $s_{ij}\equiv s_{ij}+i0$ and $m^2\equiv m^2+i0$. Expressed in these variables, the result also holds for the lower diagram. After expanding in $\varepsilon$, it agrees with the $\lambda$ expansion of the full expression for this pentagon integral given in (5.8) of~\cite{Bern:1993kr}, confirming that the leading-power contribution is fully captured by the soft-collinear region. 
The diagram where the virtual gluon is attached to the quark line \emph{after} the collinear emission (not shown in Fig.~\ref{fig:Pent2}) corresponds to a box with two massive adjacent legs. We find that two regions, the soft and the soft-collinear, fully account for the entire contribution in Euclidean kinematics.

We now analytically continue to the physical region, in which all the light-cone components are positive and $p_c^->q_c^-$. An interesting feature of the expressions for the diagrams in Fig.~\ref{fig:Pent2} are combinations that entail the cancellation of two $\mathcal{O}(\lambda)$ terms, resulting in an $\mathcal{O}(\lambda^2)$ contribution, e.g.\ for the kinematics~\eqref{eq:invariants_pentagon} belonging to the upper diagram in Fig.~\ref{fig:Pent2}, with $p_T^2\equiv -p_\perp^2 >0$,
\begin{align}
   \underbrace{s_{45}\spac s_{51}}_{\lambda} - \underbrace{m^2 s_{23}}_{\lambda} 
   = \underbrace{p_c^-\bar{p}_{\bar{c}}^{\spac+}\bigl(q_{cT}+l_{sT}\bigr)^2}_{\lambda^2}>0 \,.
\end{align}
While this has no non-trivial consequences in Euclidean kinematics, a subtlety arises upon performing the analytic continuation to physical  kinematics. To illustrate this fact, we consider the full result for the pentagon integral, which contains $\lambda$-suppressed terms with prefactor
\begin{equation}\label{eq:pentaPhase}
   \underbrace{\frac{s_{45}\spac s_{51}}{s_{45}\spac s_{51}-m^2 s_{23}}}_{\lambda^{-1}}
    \bigg[ 1 - {e^{i\pi\varepsilon\,\Theta}} 
    \bigg( 1 + \underbrace{\frac{m^2\spac s_{23}-s_{45}\spac s_{51}}{s_{45}\spac s_{51}}}_{\lambda} \bigg)^{-\varepsilon} \bigg] \,,
\end{equation}
where
\begin{equation}
   \Theta \equiv \theta(m^2) + \theta(s_{23}) - \theta(s_{45}) - \theta(s_{51}) \,.
\end{equation}
This quantity vanishes for the kinematics of the lower graph in Fig.~\ref{fig:Pent2}, and the above factor is of ${\cal O}(1)$. However, for the upper diagram $\Theta$ is non-zero, since $m^2,s_{23}>0$ and $s_{45},s_{51}<0$. This generates a non-trivial phase $e^{2i\pi\varepsilon}$, leading to a power enhancement in \eqref{eq:pentaPhase}, which compensates the power suppression and thus induces additional \emph{leading-order} terms. More generally, such terms only arise between incoming lines involving a space-like splitting and a virtual gluon attached to both the soft and the split-off gluon, as is indeed the case for the upper, but not the lower diagram in Fig.~\ref{fig:Pent2}.

Using the known results for the pentagon integrals, and subtracting the soft-collinear contribution, we can derive the extra terms in the physical scattering kinematics given in~\eqref{eq:invariants_pentagon}. In the method of regions, these terms must be generated by a new region absent in Euclidean kinematics. We find that they are proportional to $i\pi$, and are generated by a Glauber region $k\sim Q(\lambda^2,\lambda,\lambda)$. The upper pentagon in Fig.~\ref{fig:Pent2} expanded in this region takes the form (with implicit $+i0$ prescriptions)
\begin{align}
   \hspace*{-0.22cm}
   I^{\text{g}} &= i(4\pi)^{2-\varepsilon} \int\frac{d^dk}{(2\pi)^d} 
    \frac{1}{-k_{T}^2}\,\frac{1}{k^+\spac q_c^- -k_{T}^2-2 k_{T}\cdot q_{cT}}  \nonumber\\
   &\quad\times \frac{1}{\left[-k^+\spac(p_c^- -q_c^-)-q_c^+\spac p_c^- 
    - k_{T}^2-2 k_{T}\cdot q_{cT} \right]} \nonumber\\
   &\quad\times \frac{1}{\bar{p}_{\bar{c}}^{\spac +}\spac(k^- -l_{s}^-)}\,
    \frac{1}{-l_{s}^+\spac k^- -k_{T}^2+ 2k_{T}\cdot l_{sT}}
\end{align}
and is well-defined in dimensional regularization. The $k^+$ and $k^-$ integrations can be performed using residues. Evaluating the remaining two-dimensional Euclidean triangle integral, one obtains~\cite{Usyukina:1994iw}
\begin{multline}
   \hspace*{-0.22cm}
   I^{\text{g}} = - \frac{2\pi\spac i}{\bar{p}_{\bar{c}}^{\spac +}\spac p_c^-} 
    \Bigg[ \left( l_{sT}^2 + q_{cT}^2 + s_T^2 \right)
    \left( \frac{1}{\varepsilon} - \ln\big( l_{sT}^2\spac q_{cT}^2\spac s_T^2 \big) \right) \\
    + 2 l_{sT}^2 \ln l_{sT}^2 + 2 q_{cT}^2\ln q_{cT}^2 + 2 s_T^2 \ln s_T^2 \Bigg] 
    \frac{e^{-\varepsilon\gamma_E}}{l_{sT}^2\spac q_{cT}^2\spac s_T^2} \,,
\end{multline}
with $s_T=l_{sT}+q_{cT}$. Note that the triangle integral has an IR divergence even though all external momenta are off shell. This is possible because in two dimensions a single vanishing massless propagator can produce a singularity. 

It is interesting to understand the appearance of this ``hidden'' Glauber region from the representation of the pentagon integral in Feynman parameter space (more precisly the closely related Lee-Pomeransky space~\cite{Lee:2013hzt}), where variables $x_i$ are associated with the respective propagators, e.g.\ $x_1$ is linked to the propagator between vertices 1 and 2, and the ensuing $x_i$ follow in counter-clockwise direction. For the upper graph in Fig.~\ref{fig:Pent2}, the Glauber region is characterized by the scaling $(x_1,x_2,x_3,x_4,x_5)\sim (\lambda^{-2},\lambda^{-2},\lambda^{-2},\lambda^{-1},\lambda^{-2})$. For the associated $\mathcal{F}$ polynomial, one finds
\begin{align}
   \mathcal{F}
   &= -\underbrace{x_1 x_3\spac s_{23}}_{\lambda^{-3}}
    - \underbrace{x_1 x_4\spac s_{51}}_{\lambda^{-3}}
    - \underbrace{x_3 x_5\spac s_{45}}_{\lambda^{-3}} \nonumber\\
   &\quad - \underbrace{x_4 x_5\spac m^2}_{\lambda^{-3}}
    - \underbrace{x_2 x_4\spac s_{34}}_{\lambda^{-2}}
    - \underbrace{x_2 x_5\spac s_{12}}_{\lambda^{-2}} \,.
\end{align}
The first four terms constitute the leading power contribution, which using \eqref{eq:invariants_pentagon} can be factorized in the form
\begin{align}\label{eq:Landau_cancellation}
  \mathcal{F} = 
    \underbrace{\left(-q_c^-\spac x_1 + (p_c^- - q_c^-)\spac x_5\right)}_{\lambda^{-2}}
    \underbrace{\left(l_{s}^+\spac x_3-\bar{p}_{\bar{c}}^{\spac +}\spac x_4\right)}_{\lambda^{-1}} \,.
\end{align}
For physical kinematics, where all light-cone components are positive and $p_c^->q_c^-$, 
large cancellations occur within the $\mathcal{F}$ polynomial. The hidden Glauber pinch appears when both brackets in~\eqref{eq:Landau_cancellation} vanish individually, in accordance with the Landau equations~\cite{Ma:2023hrt,Gardi:2024axt}. The double cancellation with unequal coefficients of the parameters $x_i$ may be the reason why we were unable to find this region using \texttt{Asy2.1}~\cite{Pak:2010pt,Jantzen:2012mw}.

\begin{figure}
    \centering
    \includegraphics{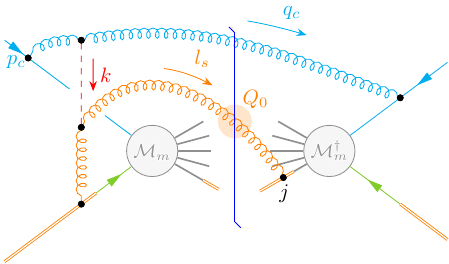}
    \caption{Example of a collinear space-like splitting with a genuine Glauber mode (red) contributing to the low-energy matrix elements. The soft gluon is emitted into the gap with constraint $Q_0$ and attaches to leg $j$ on the right-hand side. Soft Wilson lines are drawn in orange, where relevant.}
    \label{fig:QCDGlauber}
\end{figure}

With this understanding of the appearance of the Glauber region in the context of the scalar example, we now turn our attention back to the challenge at hand: explicitly verifying that also the final term in~\eqref{eq:Wm_poles} is reproduced perturbatively in the low-energy theory. With our previous discussion, we have narrowed down the class of diagrams that need to be evaluated to those involving interactions between the collinear, anti-collinear, and soft sectors such as the ones shown in Fig.~\ref{fig:Pent2}. We thus evaluate these diagrams in the soft-collinear and Glauber regions and integrate over the phase space of the real emissions. Due to the appearance of the collinear anomaly, the integration over $q_c$ is not well-defined on its own, and following~\cite{Becher:2011dz} we introduce a phase-space regulator $(\nu/q_c^-)^{2\eta}$. We find that the soft-collinear region (and the soft one in the case of the box) always leads to scaleless collinear phase-space integrals and therefore does not contribute to the cross section. This is welcome news, since the associated low-energy scale $\lambda\spac Q_0^2$ would be parametrically smaller than $Q_0^2$ and could be non-perturbative, even if $Q_0$ itself is not. What remains is the Glauber contribution. In addition to Fig.~\ref{fig:diagram1}, we also consider the mirrored diagrams in which the two incoming particles are interchanged or the Glauber exchange happens in the conjugate amplitude. The leading UV poles of these four graphs yield a contribution to the (bare) low-energy matrix element given by
\begin{align}\label{eq:Wcomp}
   \bm{\mathcal{W}}_m^{\rm bare}
   &\ni \frac{i\alpha_s^3}{12\pi^2\spac\varepsilon^3}\,
    f^{abc} f^{ade} \sum_{j>2} J_j \notag\\
   &\quad\times \bigg[ \bm{T}_{1L}^d\spac\bm{T}_{1R}^e\spac
   \bm{T}_{2L}^b\spac\bm{T}_{jR}^c \left( - \frac{1}{2\eta} - \ln\frac{\nu}{p_c^-} \right) \notag\\
   &\qquad + \bm{T}_{2L}^d\spac\bm{T}_{2R}^e\spac
   \bm{T}_{1L}^b\spac\bm{T}_{jR}^c \left( \frac{1}{2\eta} + \ln\frac{\nu\spac\bar p_{\bar c}^+}{Q_0^2} \right) \bigg] \notag\\
   &\quad - (L\leftrightarrow R) \,,
\end{align}
where the terms with $j=1,2$ have canceled out in the combination. Under the color trace with the hard function in~\eqref{hadronfact}, we can move the color generators $\bm{T}_{iL}$ to the right and replace (for $j\ne 1,2$)
\begin{equation}
   f^{ade}\,\bm{T}_{1L}^d\spac\bm{T}_{2L}^b \bm{T}_{1R}^e\spac\bm{T}_{jR}^c
   \to - \frac{i N_c}{2}\,
    \bm{T}_1^a\spac\bm{T}_2^b\spac\bm{T}_j^c \,,
\end{equation}
which leads to
\begin{align}
   \bm{\mathcal{W}}_m^{\rm bare}
   &\ni - \frac{i N_c\spac\alpha_s^3}{12\pi^2\spac\varepsilon^3}\,\bm{X}_1\spac
    \ln\frac{p_c^-\spac\bar p_{\bar c}^+}{Q_0^2} \,.
\end{align}
The divergences in $\eta$ have canceled but the associated hard logarithm remains.  It has indeed the structure required by \eqref{eq:Wm_poles} to remove the double-logarithmic part of the evolution below the scale $Q_0$. We have checked that~\eqref{eq:Wcomp} also holds for incoming gluons.

The same result can be obtained directly in SCET using the Glauber Lagrangian of \cite{Rothstein:2016bsq}. The regions analysis immediately translates into SCET diagrams such as the one shown in Fig.~\ref{fig:QCDGlauber}, where the different colors now correspond to different SCET fields and the dashed red line indicates the Glauber exchange. In the framework of~\cite{Rothstein:2016bsq}, one additionally encounters diagrams with Glauber scaling on both internal lines connecting to the soft-emission vertex. After regularizing their contribution with a Glauber regulator $|k_{g}^z|^{\eta'}$~\cite{Moult:2022lfy}, distinct from the rapidity regulator, and performing the required $0$-bin subtractions, we find that SCET reproduces \eqref{eq:Wcomp}. For our observable, the $0$-bin subtractions and the additional graph with two Glauber gluons cancel each other. We believe that it should be possible to choose a regularization scheme in which such ``non-genuine'' (or ``Cheshire'' \cite{Rothstein:2016bsq}) Glauber contributions vanish from the beginning.

In this Letter, we have uncovered a new mechanism that reconciles the breaking of collinear factorization with PDF factorization. Remarkably, it is the contribution of perturbative Glauber gluons which, in an interplay of space-like collinear splittings and soft emissions, restores the factorization of the cross section by converting double-logarithmic into single-logarithmic running at low values of the factorization scale. In the future, it will be important to understand the all-order structure of these effects, a key ingredient for the resummation of jet processes at hadron colliders to higher logarithmic accuracy \cite{Banfi:2021xzn,Becher:2023vrh,FerrarioRavasio:2023kyg} and the development of finite-$N_c$ parton showers \cite{Nagy:2007ty,Nagy:2019pjp,Forshaw:2019ver,DeAngelis:2020rvq}. This would clarify the physics of space-like collinear limits of amplitudes and pave the way to a proof of PDF factorization for a much wider class of observables.

{\em Acknowledgements --- \/} We thank Philipp B\"oer, Einan Gardi, Nicolas Schalch, Michel Stillger, Yannick Ulrich, and Xiaofeng Xu for useful discussions, and are grateful to Yao Ma for help and for sharing his insight into the analysis in Feynman parameter space. This work was supported by the Swiss National Science Foundation (SNF) under grant 200021\_219377, by the European Research Council (ERC) under the European Union’s Horizon 2022 Research and Innovation Program (ERC Advanced Grant agreement No.~101097780, EFT4jets), and by the Cluster of Excellence PRISMA+ (Precision Physics, Fundamental Interactions, and Structure of Matter, EXC 2118/1) funded by the German Research Foundation (DFG) under Germany’s Excellence Strategy (Project ID 390831469).  

\bibliography{Glauber}

\begin{thebibliography}{52}%
\makeatletter
\providecommand \@ifxundefined [1]{%
 \@ifx{#1\undefined}
}%
\providecommand \@ifnum [1]{%
 \ifnum #1\expandafter \@firstoftwo
 \else \expandafter \@secondoftwo
 \fi
}%
\providecommand \@ifx [1]{%
 \ifx #1\expandafter \@firstoftwo
 \else \expandafter \@secondoftwo
 \fi
}%
\providecommand \natexlab [1]{#1}%
\providecommand \enquote  [1]{``#1''}%
\providecommand \bibnamefont  [1]{#1}%
\providecommand \bibfnamefont [1]{#1}%
\providecommand \citenamefont [1]{#1}%
\providecommand \href@noop [0]{\@secondoftwo}%
\providecommand \href [0]{\begingroup \@sanitize@url \@href}%
\providecommand \@href[1]{\@@startlink{#1}\@@href}%
\providecommand \@@href[1]{\endgroup#1\@@endlink}%
\providecommand \@sanitize@url [0]{\catcode `\\12\catcode `\$12\catcode `\&12\catcode `\#12\catcode `\^12\catcode `\_12\catcode `\%12\relax}%
\providecommand \@@startlink[1]{}%
\providecommand \@@endlink[0]{}%
\providecommand \url  [0]{\begingroup\@sanitize@url \@url }%
\providecommand \@url [1]{\endgroup\@href {#1}{\urlprefix }}%
\providecommand \urlprefix  [0]{URL }%
\providecommand \Eprint [0]{\href }%
\providecommand \doibase [0]{https://doi.org/}%
\providecommand \selectlanguage [0]{\@gobble}%
\providecommand \bibinfo  [0]{\@secondoftwo}%
\providecommand \bibfield  [0]{\@secondoftwo}%
\providecommand \translation [1]{[#1]}%
\providecommand \BibitemOpen [0]{}%
\providecommand \bibitemStop [0]{}%
\providecommand \bibitemNoStop [0]{.\EOS\space}%
\providecommand \EOS [0]{\spacefactor3000\relax}%
\providecommand \BibitemShut  [1]{\csname bibitem#1\endcsname}%
\let\auto@bib@innerbib\@empty
\bibitem [{\citenamefont {Collins}\ \emph {et~al.}(1985)\citenamefont {Collins}, \citenamefont {Soper},\ and\ \citenamefont {Sterman}}]{Collins:1985ue}%
  \BibitemOpen
  \bibfield  {author} {\bibinfo {author} {\bibfnamefont {J.~C.}\ \bibnamefont {Collins}}, \bibinfo {author} {\bibfnamefont {D.~E.}\ \bibnamefont {Soper}},\ and\ \bibinfo {author} {\bibfnamefont {G.~F.}\ \bibnamefont {Sterman}},\ }\bibfield  {title} {\bibinfo {title} {{Factorization for Short Distance Hadron - Hadron Scattering}},\ }\href {https://doi.org/10.1016/0550-3213(85)90565-6} {\bibfield  {journal} {\bibinfo  {journal} {Nucl. Phys. B}\ }\textbf {\bibinfo {volume} {261}},\ \bibinfo {pages} {104} (\bibinfo {year} {1985})}\BibitemShut {NoStop}%
\bibitem [{\citenamefont {Collins}\ and\ \citenamefont {Qiu}(2007)}]{Collins:2007nk}%
  \BibitemOpen
  \bibfield  {author} {\bibinfo {author} {\bibfnamefont {J.}~\bibnamefont {Collins}}\ and\ \bibinfo {author} {\bibfnamefont {J.-W.}\ \bibnamefont {Qiu}},\ }\bibfield  {title} {\bibinfo {title} {{$k_{T}$ factorization is violated in production of high-transverse-momentum particles in hadron-hadron collisions}},\ }\href {https://doi.org/10.1103/PhysRevD.75.114014} {\bibfield  {journal} {\bibinfo  {journal} {Phys. Rev. D}\ }\textbf {\bibinfo {volume} {75}},\ \bibinfo {pages} {114014} (\bibinfo {year} {2007})},\ \Eprint {https://arxiv.org/abs/0705.2141} {arXiv:0705.2141 [hep-ph]} \BibitemShut {NoStop}%
\bibitem [{\citenamefont {Gaunt}(2014)}]{Gaunt:2014ska}%
  \BibitemOpen
  \bibfield  {author} {\bibinfo {author} {\bibfnamefont {J.~R.}\ \bibnamefont {Gaunt}},\ }\bibfield  {title} {\bibinfo {title} {{Glauber Gluons and Multiple Parton Interactions}},\ }\href {https://doi.org/10.1007/JHEP07(2014)110} {\bibfield  {journal} {\bibinfo  {journal} {JHEP}\ }\textbf {\bibinfo {volume} {07}},\ \bibinfo {pages} {110}},\ \Eprint {https://arxiv.org/abs/1405.2080} {arXiv:1405.2080 [hep-ph]} \BibitemShut {NoStop}%
\bibitem [{\citenamefont {Zeng}(2015)}]{Zeng:2015iba}%
  \BibitemOpen
  \bibfield  {author} {\bibinfo {author} {\bibfnamefont {M.}~\bibnamefont {Zeng}},\ }\bibfield  {title} {\bibinfo {title} {{Drell-Yan process with jet vetoes: breaking of generalized factorization}},\ }\href {https://doi.org/10.1007/JHEP10(2015)189} {\bibfield  {journal} {\bibinfo  {journal} {JHEP}\ }\textbf {\bibinfo {volume} {10}},\ \bibinfo {pages} {189}},\ \Eprint {https://arxiv.org/abs/1507.01652} {arXiv:1507.01652 [hep-ph]} \BibitemShut {NoStop}%
\bibitem [{\citenamefont {Catani}\ \emph {et~al.}(2012)\citenamefont {Catani}, \citenamefont {de~Florian},\ and\ \citenamefont {Rodrigo}}]{Catani:2011st}%
  \BibitemOpen
  \bibfield  {author} {\bibinfo {author} {\bibfnamefont {S.}~\bibnamefont {Catani}}, \bibinfo {author} {\bibfnamefont {D.}~\bibnamefont {de~Florian}},\ and\ \bibinfo {author} {\bibfnamefont {G.}~\bibnamefont {Rodrigo}},\ }\bibfield  {title} {\bibinfo {title} {{Space-like (versus time-like) collinear limits in QCD: Is factorization violated?}},\ }\href {https://doi.org/10.1007/JHEP07(2012)026} {\bibfield  {journal} {\bibinfo  {journal} {JHEP}\ }\textbf {\bibinfo {volume} {07}},\ \bibinfo {pages} {026}},\ \Eprint {https://arxiv.org/abs/1112.4405} {arXiv:1112.4405 [hep-ph]} \BibitemShut {NoStop}%
\bibitem [{\citenamefont {Forshaw}\ \emph {et~al.}(2012)\citenamefont {Forshaw}, \citenamefont {Seymour},\ and\ \citenamefont {Siodmok}}]{Forshaw:2012bi}%
  \BibitemOpen
  \bibfield  {author} {\bibinfo {author} {\bibfnamefont {J.~R.}\ \bibnamefont {Forshaw}}, \bibinfo {author} {\bibfnamefont {M.~H.}\ \bibnamefont {Seymour}},\ and\ \bibinfo {author} {\bibfnamefont {A.}~\bibnamefont {Siodmok}},\ }\bibfield  {title} {\bibinfo {title} {{On the Breaking of Collinear Factorization in QCD}},\ }\href {https://doi.org/10.1007/JHEP11(2012)066} {\bibfield  {journal} {\bibinfo  {journal} {JHEP}\ }\textbf {\bibinfo {volume} {11}},\ \bibinfo {pages} {066}},\ \Eprint {https://arxiv.org/abs/1206.6363} {arXiv:1206.6363 [hep-ph]} \BibitemShut {NoStop}%
\bibitem [{\citenamefont {Schwartz}\ \emph {et~al.}(2017)\citenamefont {Schwartz}, \citenamefont {Yan},\ and\ \citenamefont {Zhu}}]{Schwartz:2017nmr}%
  \BibitemOpen
  \bibfield  {author} {\bibinfo {author} {\bibfnamefont {M.~D.}\ \bibnamefont {Schwartz}}, \bibinfo {author} {\bibfnamefont {K.}~\bibnamefont {Yan}},\ and\ \bibinfo {author} {\bibfnamefont {H.~X.}\ \bibnamefont {Zhu}},\ }\bibfield  {title} {\bibinfo {title} {{Collinear factorization violation and effective field theory}},\ }\href {https://doi.org/10.1103/PhysRevD.96.056005} {\bibfield  {journal} {\bibinfo  {journal} {Phys. Rev. D}\ }\textbf {\bibinfo {volume} {96}},\ \bibinfo {pages} {056005} (\bibinfo {year} {2017})},\ \Eprint {https://arxiv.org/abs/1703.08572} {arXiv:1703.08572 [hep-ph]} \BibitemShut {NoStop}%
\bibitem [{\citenamefont {Cieri}\ \emph {et~al.}(2024)\citenamefont {Cieri}, \citenamefont {Dhani},\ and\ \citenamefont {Rodrigo}}]{Cieri:2024ytf}%
  \BibitemOpen
  \bibfield  {author} {\bibinfo {author} {\bibfnamefont {L.}~\bibnamefont {Cieri}}, \bibinfo {author} {\bibfnamefont {P.~K.}\ \bibnamefont {Dhani}},\ and\ \bibinfo {author} {\bibfnamefont {G.}~\bibnamefont {Rodrigo}},\ }\bibfield  {title} {\bibinfo {title} {{Catani's generalization of collinear factorization breaking}},\ }\href@noop {} {\  (\bibinfo {year} {2024})},\ \Eprint {https://arxiv.org/abs/2402.14749} {arXiv:2402.14749 [hep-ph]} \BibitemShut {NoStop}%
\bibitem [{\citenamefont {Henn}\ \emph {et~al.}(2024)\citenamefont {Henn}, \citenamefont {Ma}, \citenamefont {Xu}, \citenamefont {Yan}, \citenamefont {Zhang},\ and\ \citenamefont {Zhu}}]{Henn:2024qjq}%
  \BibitemOpen
  \bibfield  {author} {\bibinfo {author} {\bibfnamefont {J.}~\bibnamefont {Henn}}, \bibinfo {author} {\bibfnamefont {R.}~\bibnamefont {Ma}}, \bibinfo {author} {\bibfnamefont {Y.}~\bibnamefont {Xu}}, \bibinfo {author} {\bibfnamefont {K.}~\bibnamefont {Yan}}, \bibinfo {author} {\bibfnamefont {Y.}~\bibnamefont {Zhang}},\ and\ \bibinfo {author} {\bibfnamefont {H.~X.}\ \bibnamefont {Zhu}},\ }\bibfield  {title} {\bibinfo {title} {{Two-Loop Spacelike Splitting Amplitude for N=4 Super-Yang-Mills Theory}},\ }\href@noop {} {\  (\bibinfo {year} {2024})},\ \Eprint {https://arxiv.org/abs/2406.14604} {arXiv:2406.14604 [hep-th]} \BibitemShut {NoStop}%
\bibitem [{\citenamefont {Guan}\ \emph {et~al.}(2024)\citenamefont {Guan}, \citenamefont {Herzog}, \citenamefont {Ma}, \citenamefont {Mistlberger},\ and\ \citenamefont {Suresh}}]{Guan:2024hlf}%
  \BibitemOpen
  \bibfield  {author} {\bibinfo {author} {\bibfnamefont {X.}~\bibnamefont {Guan}}, \bibinfo {author} {\bibfnamefont {F.}~\bibnamefont {Herzog}}, \bibinfo {author} {\bibfnamefont {Y.}~\bibnamefont {Ma}}, \bibinfo {author} {\bibfnamefont {B.}~\bibnamefont {Mistlberger}},\ and\ \bibinfo {author} {\bibfnamefont {A.}~\bibnamefont {Suresh}},\ }\bibfield  {title} {\bibinfo {title} {{Splitting amplitudes at N$^3$LO in QCD}},\ }\href@noop {} {\  (\bibinfo {year} {2024})},\ \Eprint {https://arxiv.org/abs/2408.03019} {arXiv:2408.03019 [hep-ph]} \BibitemShut {NoStop}%
\bibitem [{\citenamefont {Forshaw}\ \emph {et~al.}(2008)\citenamefont {Forshaw}, \citenamefont {Kyrieleis},\ and\ \citenamefont {Seymour}}]{Forshaw:2008cq}%
  \BibitemOpen
  \bibfield  {author} {\bibinfo {author} {\bibfnamefont {J.~R.}\ \bibnamefont {Forshaw}}, \bibinfo {author} {\bibfnamefont {A.}~\bibnamefont {Kyrieleis}},\ and\ \bibinfo {author} {\bibfnamefont {M.~H.}\ \bibnamefont {Seymour}},\ }\bibfield  {title} {\bibinfo {title} {{Super-leading logarithms in non-global observables in QCD: Colour basis independent calculation}},\ }\href {https://doi.org/10.1088/1126-6708/2008/09/128} {\bibfield  {journal} {\bibinfo  {journal} {JHEP}\ }\textbf {\bibinfo {volume} {09}},\ \bibinfo {pages} {128}},\ \Eprint {https://arxiv.org/abs/0808.1269} {arXiv:0808.1269 [hep-ph]} \BibitemShut {NoStop}%
\bibitem [{\citenamefont {Becher}\ \emph {et~al.}(2021)\citenamefont {Becher}, \citenamefont {Neubert},\ and\ \citenamefont {Shao}}]{Becher:2021zkk}%
  \BibitemOpen
  \bibfield  {author} {\bibinfo {author} {\bibfnamefont {T.}~\bibnamefont {Becher}}, \bibinfo {author} {\bibfnamefont {M.}~\bibnamefont {Neubert}},\ and\ \bibinfo {author} {\bibfnamefont {D.~Y.}\ \bibnamefont {Shao}},\ }\bibfield  {title} {\bibinfo {title} {{Resummation of Super-Leading Logarithms}},\ }\href {https://doi.org/10.1103/PhysRevLett.127.212002} {\bibfield  {journal} {\bibinfo  {journal} {Phys. Rev. Lett.}\ }\textbf {\bibinfo {volume} {127}},\ \bibinfo {pages} {212002} (\bibinfo {year} {2021})},\ \Eprint {https://arxiv.org/abs/2107.01212} {arXiv:2107.01212 [hep-ph]} \BibitemShut {NoStop}%
\bibitem [{\citenamefont {Becher}\ \emph {et~al.}(2023)\citenamefont {Becher}, \citenamefont {Neubert}, \citenamefont {Shao},\ and\ \citenamefont {Stillger}}]{Becher:2023mtx}%
  \BibitemOpen
  \bibfield  {author} {\bibinfo {author} {\bibfnamefont {T.}~\bibnamefont {Becher}}, \bibinfo {author} {\bibfnamefont {M.}~\bibnamefont {Neubert}}, \bibinfo {author} {\bibfnamefont {D.~Y.}\ \bibnamefont {Shao}},\ and\ \bibinfo {author} {\bibfnamefont {M.}~\bibnamefont {Stillger}},\ }\bibfield  {title} {\bibinfo {title} {{Factorization of non-global LHC observables and resummation of super-leading logarithms}},\ }\href {https://doi.org/10.1007/JHEP12(2023)116} {\bibfield  {journal} {\bibinfo  {journal} {JHEP}\ }\textbf {\bibinfo {volume} {12}},\ \bibinfo {pages} {116}},\ \Eprint {https://arxiv.org/abs/2307.06359} {arXiv:2307.06359 [hep-ph]} \BibitemShut {NoStop}%
\bibitem [{\citenamefont {B\"oer}\ \emph {et~al.}(2024)\citenamefont {B\"oer}, \citenamefont {Hager}, \citenamefont {Neubert}, \citenamefont {Stillger},\ and\ \citenamefont {Xu}}]{Boer:2024hzh}%
  \BibitemOpen
  \bibfield  {author} {\bibinfo {author} {\bibfnamefont {P.}~\bibnamefont {B\"oer}}, \bibinfo {author} {\bibfnamefont {P.}~\bibnamefont {Hager}}, \bibinfo {author} {\bibfnamefont {M.}~\bibnamefont {Neubert}}, \bibinfo {author} {\bibfnamefont {M.}~\bibnamefont {Stillger}},\ and\ \bibinfo {author} {\bibfnamefont {X.}~\bibnamefont {Xu}},\ }\bibfield  {title} {\bibinfo {title} {{Renormalization-group improved resummation of super-leading logarithms}},\ }\href {https://doi.org/10.1007/JHEP08(2024)035} {\bibfield  {journal} {\bibinfo  {journal} {JHEP}\ }\textbf {\bibinfo {volume} {08}},\ \bibinfo {pages} {035}},\ \Eprint {https://arxiv.org/abs/2405.05305} {arXiv:2405.05305 [hep-ph]} \BibitemShut {NoStop}%
\bibitem [{\citenamefont {Bauer}\ \emph {et~al.}(2002{\natexlab{a}})\citenamefont {Bauer}, \citenamefont {Pirjol},\ and\ \citenamefont {Stewart}}]{Bauer:2001yt}%
  \BibitemOpen
  \bibfield  {author} {\bibinfo {author} {\bibfnamefont {C.~W.}\ \bibnamefont {Bauer}}, \bibinfo {author} {\bibfnamefont {D.}~\bibnamefont {Pirjol}},\ and\ \bibinfo {author} {\bibfnamefont {I.~W.}\ \bibnamefont {Stewart}},\ }\bibfield  {title} {\bibinfo {title} {{Soft collinear factorization in effective field theory}},\ }\href {https://doi.org/10.1103/PhysRevD.65.054022} {\bibfield  {journal} {\bibinfo  {journal} {Phys. Rev. D}\ }\textbf {\bibinfo {volume} {65}},\ \bibinfo {pages} {054022} (\bibinfo {year} {2002}{\natexlab{a}})},\ \Eprint {https://arxiv.org/abs/hep-ph/0109045} {arXiv:hep-ph/0109045} \BibitemShut {NoStop}%
\bibitem [{\citenamefont {Bauer}\ \emph {et~al.}(2002{\natexlab{b}})\citenamefont {Bauer}, \citenamefont {Fleming}, \citenamefont {Pirjol}, \citenamefont {Rothstein},\ and\ \citenamefont {Stewart}}]{Bauer:2002nz}%
  \BibitemOpen
  \bibfield  {author} {\bibinfo {author} {\bibfnamefont {C.~W.}\ \bibnamefont {Bauer}}, \bibinfo {author} {\bibfnamefont {S.}~\bibnamefont {Fleming}}, \bibinfo {author} {\bibfnamefont {D.}~\bibnamefont {Pirjol}}, \bibinfo {author} {\bibfnamefont {I.~Z.}\ \bibnamefont {Rothstein}},\ and\ \bibinfo {author} {\bibfnamefont {I.~W.}\ \bibnamefont {Stewart}},\ }\bibfield  {title} {\bibinfo {title} {{Hard scattering factorization from effective field theory}},\ }\href {https://doi.org/10.1103/PhysRevD.66.014017} {\bibfield  {journal} {\bibinfo  {journal} {Phys. Rev. D}\ }\textbf {\bibinfo {volume} {66}},\ \bibinfo {pages} {014017} (\bibinfo {year} {2002}{\natexlab{b}})},\ \Eprint {https://arxiv.org/abs/hep-ph/0202088} {arXiv:hep-ph/0202088} \BibitemShut {NoStop}%
\bibitem [{\citenamefont {Beneke}\ \emph {et~al.}(2002)\citenamefont {Beneke}, \citenamefont {Chapovsky}, \citenamefont {Diehl},\ and\ \citenamefont {Feldmann}}]{Beneke:2002ph}%
  \BibitemOpen
  \bibfield  {author} {\bibinfo {author} {\bibfnamefont {M.}~\bibnamefont {Beneke}}, \bibinfo {author} {\bibfnamefont {A.~P.}\ \bibnamefont {Chapovsky}}, \bibinfo {author} {\bibfnamefont {M.}~\bibnamefont {Diehl}},\ and\ \bibinfo {author} {\bibfnamefont {T.}~\bibnamefont {Feldmann}},\ }\bibfield  {title} {\bibinfo {title} {{Soft collinear effective theory and heavy to light currents beyond leading power}},\ }\href {https://doi.org/10.1016/S0550-3213(02)00687-9} {\bibfield  {journal} {\bibinfo  {journal} {Nucl. Phys. B}\ }\textbf {\bibinfo {volume} {643}},\ \bibinfo {pages} {431} (\bibinfo {year} {2002})},\ \Eprint {https://arxiv.org/abs/hep-ph/0206152} {arXiv:hep-ph/0206152} \BibitemShut {NoStop}%
\bibitem [{\citenamefont {Beneke}\ and\ \citenamefont {Feldmann}(2003)}]{Beneke:2002ni}%
  \BibitemOpen
  \bibfield  {author} {\bibinfo {author} {\bibfnamefont {M.}~\bibnamefont {Beneke}}\ and\ \bibinfo {author} {\bibfnamefont {T.}~\bibnamefont {Feldmann}},\ }\bibfield  {title} {\bibinfo {title} {{Multipole expanded soft collinear effective theory with nonAbelian gauge symmetry}},\ }\href {https://doi.org/10.1016/S0370-2693(02)03204-5} {\bibfield  {journal} {\bibinfo  {journal} {Phys. Lett. B}\ }\textbf {\bibinfo {volume} {553}},\ \bibinfo {pages} {267} (\bibinfo {year} {2003})},\ \Eprint {https://arxiv.org/abs/hep-ph/0211358} {arXiv:hep-ph/0211358} \BibitemShut {NoStop}%
\bibitem [{\citenamefont {Beneke}\ and\ \citenamefont {Smirnov}(1998)}]{Beneke:1997zp}%
  \BibitemOpen
  \bibfield  {author} {\bibinfo {author} {\bibfnamefont {M.}~\bibnamefont {Beneke}}\ and\ \bibinfo {author} {\bibfnamefont {V.~A.}\ \bibnamefont {Smirnov}},\ }\bibfield  {title} {\bibinfo {title} {{Asymptotic expansion of Feynman integrals near threshold}},\ }\href {https://doi.org/10.1016/S0550-3213(98)00138-2} {\bibfield  {journal} {\bibinfo  {journal} {Nucl. Phys.}\ }\textbf {\bibinfo {volume} {B522}},\ \bibinfo {pages} {321} (\bibinfo {year} {1998})},\ \Eprint {https://arxiv.org/abs/hep-ph/9711391} {arXiv:hep-ph/9711391 [hep-ph]} \BibitemShut {NoStop}%
\bibitem [{\citenamefont {Smirnov}(2002)}]{Smirnov:2002pj}%
  \BibitemOpen
  \bibfield  {author} {\bibinfo {author} {\bibfnamefont {V.~A.}\ \bibnamefont {Smirnov}},\ }\bibfield  {title} {\bibinfo {title} {{Applied asymptotic expansions in momenta and masses}},\ }\href@noop {} {\bibfield  {journal} {\bibinfo  {journal} {Springer Tracts Mod. Phys.}\ }\textbf {\bibinfo {volume} {177}},\ \bibinfo {pages} {1} (\bibinfo {year} {2002})}\BibitemShut {NoStop}%
\bibitem [{\citenamefont {Jantzen}\ \emph {et~al.}(2012)\citenamefont {Jantzen}, \citenamefont {Smirnov},\ and\ \citenamefont {Smirnov}}]{Jantzen:2012mw}%
  \BibitemOpen
  \bibfield  {author} {\bibinfo {author} {\bibfnamefont {B.}~\bibnamefont {Jantzen}}, \bibinfo {author} {\bibfnamefont {A.~V.}\ \bibnamefont {Smirnov}},\ and\ \bibinfo {author} {\bibfnamefont {V.~A.}\ \bibnamefont {Smirnov}},\ }\bibfield  {title} {\bibinfo {title} {{Expansion by regions: revealing potential and Glauber regions automatically}},\ }\href {https://doi.org/10.1140/epjc/s10052-012-2139-2} {\bibfield  {journal} {\bibinfo  {journal} {Eur. Phys. J. C}\ }\textbf {\bibinfo {volume} {72}},\ \bibinfo {pages} {2139} (\bibinfo {year} {2012})},\ \Eprint {https://arxiv.org/abs/1206.0546} {arXiv:1206.0546 [hep-ph]} \BibitemShut {NoStop}%
\bibitem [{\citenamefont {Ananthanarayan}\ \emph {et~al.}(2020)\citenamefont {Ananthanarayan}, \citenamefont {Das},\ and\ \citenamefont {Sarkar}}]{Ananthanarayan:2020ptw}%
  \BibitemOpen
  \bibfield  {author} {\bibinfo {author} {\bibfnamefont {B.}~\bibnamefont {Ananthanarayan}}, \bibinfo {author} {\bibfnamefont {A.~B.}\ \bibnamefont {Das}},\ and\ \bibinfo {author} {\bibfnamefont {R.}~\bibnamefont {Sarkar}},\ }\bibfield  {title} {\bibinfo {title} {{Asymptotic analysis of Feynman diagrams and their maximal cuts}},\ }\href {https://doi.org/10.1140/epjc/s10052-020-08609-0} {\bibfield  {journal} {\bibinfo  {journal} {Eur. Phys. J. C}\ }\textbf {\bibinfo {volume} {80}},\ \bibinfo {pages} {1131} (\bibinfo {year} {2020})},\ \Eprint {https://arxiv.org/abs/2003.02451} {arXiv:2003.02451 [hep-ph]} \BibitemShut {NoStop}%
\bibitem [{\citenamefont {Gardi}\ \emph {et~al.}(2023)\citenamefont {Gardi}, \citenamefont {Herzog}, \citenamefont {Jones}, \citenamefont {Ma},\ and\ \citenamefont {Schlenk}}]{Gardi:2022khw}%
  \BibitemOpen
  \bibfield  {author} {\bibinfo {author} {\bibfnamefont {E.}~\bibnamefont {Gardi}}, \bibinfo {author} {\bibfnamefont {F.}~\bibnamefont {Herzog}}, \bibinfo {author} {\bibfnamefont {S.}~\bibnamefont {Jones}}, \bibinfo {author} {\bibfnamefont {Y.}~\bibnamefont {Ma}},\ and\ \bibinfo {author} {\bibfnamefont {J.}~\bibnamefont {Schlenk}},\ }\bibfield  {title} {\bibinfo {title} {{The on-shell expansion: from Landau equations to the Newton polytope}},\ }\href {https://doi.org/10.1007/JHEP07(2023)197} {\bibfield  {journal} {\bibinfo  {journal} {JHEP}\ }\textbf {\bibinfo {volume} {07}},\ \bibinfo {pages} {197}},\ \Eprint {https://arxiv.org/abs/2211.14845} {arXiv:2211.14845 [hep-th]} \BibitemShut {NoStop}%
\bibitem [{\citenamefont {Beneke}\ \emph {et~al.}(2024)\citenamefont {Beneke}, \citenamefont {Hager},\ and\ \citenamefont {Sanfilippo}}]{Beneke:2023wmt}%
  \BibitemOpen
  \bibfield  {author} {\bibinfo {author} {\bibfnamefont {M.}~\bibnamefont {Beneke}}, \bibinfo {author} {\bibfnamefont {P.}~\bibnamefont {Hager}},\ and\ \bibinfo {author} {\bibfnamefont {A.~F.}\ \bibnamefont {Sanfilippo}},\ }\bibfield  {title} {\bibinfo {title} {{Cosmological correlators in massless \ensuremath{\phi}$^{4}$-theory and the method of regions}},\ }\href {https://doi.org/10.1007/JHEP04(2024)006} {\bibfield  {journal} {\bibinfo  {journal} {JHEP}\ }\textbf {\bibinfo {volume} {04}},\ \bibinfo {pages} {006}},\ \Eprint {https://arxiv.org/abs/2312.06766} {arXiv:2312.06766 [hep-th]} \BibitemShut {NoStop}%
\bibitem [{\citenamefont {Ma}(2023)}]{Ma:2023hrt}%
  \BibitemOpen
  \bibfield  {author} {\bibinfo {author} {\bibfnamefont {Y.}~\bibnamefont {Ma}},\ }\bibfield  {title} {\bibinfo {title} {{Identifying regions in wide-angle scattering via graph-theoretical approaches}},\ }\href@noop {} {\  (\bibinfo {year} {2023})},\ \Eprint {https://arxiv.org/abs/2312.14012} {arXiv:2312.14012 [hep-ph]} \BibitemShut {NoStop}%
\bibitem [{\citenamefont {Smirnov}\ and\ \citenamefont {Wunder}(2024)}]{Smirnov:2024pbj}%
  \BibitemOpen
  \bibfield  {author} {\bibinfo {author} {\bibfnamefont {V.~A.}\ \bibnamefont {Smirnov}}\ and\ \bibinfo {author} {\bibfnamefont {F.}~\bibnamefont {Wunder}},\ }\bibfield  {title} {\bibinfo {title} {{Expansion by regions meets angular integrals}},\ }\href@noop {} {\  (\bibinfo {year} {2024})},\ \Eprint {https://arxiv.org/abs/2405.13120} {arXiv:2405.13120 [hep-ph]} \BibitemShut {NoStop}%
\bibitem [{\citenamefont {Gardi}\ \emph {et~al.}(2024)\citenamefont {Gardi}, \citenamefont {Herzog}, \citenamefont {Jones},\ and\ \citenamefont {Ma}}]{Gardi:2024axt}%
  \BibitemOpen
  \bibfield  {author} {\bibinfo {author} {\bibfnamefont {E.}~\bibnamefont {Gardi}}, \bibinfo {author} {\bibfnamefont {F.}~\bibnamefont {Herzog}}, \bibinfo {author} {\bibfnamefont {S.}~\bibnamefont {Jones}},\ and\ \bibinfo {author} {\bibfnamefont {Y.}~\bibnamefont {Ma}},\ }\bibfield  {title} {\bibinfo {title} {{Dissecting polytopes: Landau singularities and asymptotic expansions in $2\to 2$ scattering}},\ }\href@noop {} {\  (\bibinfo {year} {2024})},\ \Eprint {https://arxiv.org/abs/2407.13738} {arXiv:2407.13738 [hep-th]} \BibitemShut {NoStop}%
\bibitem [{\citenamefont {Aad}\ \emph {et~al.}(2011)\citenamefont {Aad} \emph {et~al.}}]{ATLAS:2011yyh}%
  \BibitemOpen
  \bibfield  {author} {\bibinfo {author} {\bibfnamefont {G.}~\bibnamefont {Aad}} \emph {et~al.} (\bibinfo {collaboration} {ATLAS}),\ }\bibfield  {title} {\bibinfo {title} {{Measurement of dijet production with a veto on additional central jet activity in $pp$ collisions at $\sqrt{s}=7$ TeV using the ATLAS detector}},\ }\href {https://doi.org/10.1007/JHEP09(2011)053} {\bibfield  {journal} {\bibinfo  {journal} {JHEP}\ }\textbf {\bibinfo {volume} {09}},\ \bibinfo {pages} {053}},\ \Eprint {https://arxiv.org/abs/1107.1641} {arXiv:1107.1641 [hep-ex]} \BibitemShut {NoStop}%
\bibitem [{\citenamefont {Balsiger}\ \emph {et~al.}(2018)\citenamefont {Balsiger}, \citenamefont {Becher},\ and\ \citenamefont {Shao}}]{Balsiger:2018ezi}%
  \BibitemOpen
  \bibfield  {author} {\bibinfo {author} {\bibfnamefont {M.}~\bibnamefont {Balsiger}}, \bibinfo {author} {\bibfnamefont {T.}~\bibnamefont {Becher}},\ and\ \bibinfo {author} {\bibfnamefont {D.~Y.}\ \bibnamefont {Shao}},\ }\bibfield  {title} {\bibinfo {title} {{Non-global logarithms in jet and isolation cone cross sections}},\ }\href {https://doi.org/10.1007/JHEP08(2018)104} {\bibfield  {journal} {\bibinfo  {journal} {JHEP}\ }\textbf {\bibinfo {volume} {08}},\ \bibinfo {pages} {104}},\ \Eprint {https://arxiv.org/abs/1803.07045} {arXiv:1803.07045 [hep-ph]} \BibitemShut {NoStop}%
\bibitem [{Note1()}]{Note1}%
  \BibitemOpen
  \bibinfo {note} {Note that the normalization of $\protect \bm {\Gamma }^c$ and $\protect \bm {V}^G$ differs from the one used in~\cite {Becher:2021zkk,Becher:2023mtx} by a factor 4.}\BibitemShut {Stop}%
\bibitem [{\citenamefont {Becher}\ and\ \citenamefont {Neubert}(2009)}]{Becher:2009qa}%
  \BibitemOpen
  \bibfield  {author} {\bibinfo {author} {\bibfnamefont {T.}~\bibnamefont {Becher}}\ and\ \bibinfo {author} {\bibfnamefont {M.}~\bibnamefont {Neubert}},\ }\bibfield  {title} {\bibinfo {title} {{On the Structure of Infrared Singularities of Gauge-Theory Amplitudes}},\ }\href {https://doi.org/10.1088/1126-6708/2009/06/081} {\bibfield  {journal} {\bibinfo  {journal} {JHEP}\ }\textbf {\bibinfo {volume} {06}},\ \bibinfo {pages} {081}},\ \bibinfo {note} {[Erratum: JHEP 11, 024 (2013)]},\ \Eprint {https://arxiv.org/abs/0903.1126} {arXiv:0903.1126 [hep-ph]} \BibitemShut {NoStop}%
\bibitem [{\citenamefont {Catani}\ and\ \citenamefont {Grazzini}(2000)}]{Catani:2000pi}%
  \BibitemOpen
  \bibfield  {author} {\bibinfo {author} {\bibfnamefont {S.}~\bibnamefont {Catani}}\ and\ \bibinfo {author} {\bibfnamefont {M.}~\bibnamefont {Grazzini}},\ }\bibfield  {title} {\bibinfo {title} {{The soft gluon current at one loop order}},\ }\href {https://doi.org/10.1016/S0550-3213(00)00572-1} {\bibfield  {journal} {\bibinfo  {journal} {Nucl. Phys. B}\ }\textbf {\bibinfo {volume} {591}},\ \bibinfo {pages} {435} (\bibinfo {year} {2000})},\ \Eprint {https://arxiv.org/abs/hep-ph/0007142} {arXiv:hep-ph/0007142} \BibitemShut {NoStop}%
\bibitem [{\citenamefont {Duhr}\ and\ \citenamefont {Gehrmann}(2013)}]{Duhr:2013msa}%
  \BibitemOpen
  \bibfield  {author} {\bibinfo {author} {\bibfnamefont {C.}~\bibnamefont {Duhr}}\ and\ \bibinfo {author} {\bibfnamefont {T.}~\bibnamefont {Gehrmann}},\ }\bibfield  {title} {\bibinfo {title} {{The two-loop soft current in dimensional regularization}},\ }\href {https://doi.org/10.1016/j.physletb.2013.10.063} {\bibfield  {journal} {\bibinfo  {journal} {Phys. Lett. B}\ }\textbf {\bibinfo {volume} {727}},\ \bibinfo {pages} {452} (\bibinfo {year} {2013})},\ \Eprint {https://arxiv.org/abs/1309.4393} {arXiv:1309.4393 [hep-ph]} \BibitemShut {NoStop}%
\bibitem [{\citenamefont {Dixon}\ \emph {et~al.}(2020)\citenamefont {Dixon}, \citenamefont {Herrmann}, \citenamefont {Yan},\ and\ \citenamefont {Zhu}}]{Dixon:2019lnw}%
  \BibitemOpen
  \bibfield  {author} {\bibinfo {author} {\bibfnamefont {L.~J.}\ \bibnamefont {Dixon}}, \bibinfo {author} {\bibfnamefont {E.}~\bibnamefont {Herrmann}}, \bibinfo {author} {\bibfnamefont {K.}~\bibnamefont {Yan}},\ and\ \bibinfo {author} {\bibfnamefont {H.~X.}\ \bibnamefont {Zhu}},\ }\bibfield  {title} {\bibinfo {title} {{Soft gluon emission at two loops in full color}},\ }\href {https://doi.org/10.1007/JHEP05(2020)135} {\bibfield  {journal} {\bibinfo  {journal} {JHEP}\ }\textbf {\bibinfo {volume} {05}},\ \bibinfo {pages} {135}},\ \bibinfo {note} {[Erratum: JHEP 06, 143 (2024)]},\ \Eprint {https://arxiv.org/abs/1912.09370} {arXiv:1912.09370 [hep-ph]} \BibitemShut {NoStop}%
\bibitem [{\citenamefont {Becher}\ and\ \citenamefont {Neubert}(2011)}]{Becher:2010tm}%
  \BibitemOpen
  \bibfield  {author} {\bibinfo {author} {\bibfnamefont {T.}~\bibnamefont {Becher}}\ and\ \bibinfo {author} {\bibfnamefont {M.}~\bibnamefont {Neubert}},\ }\bibfield  {title} {\bibinfo {title} {{Drell-Yan Production at Small $q_T$, Transverse Parton Distributions and the Collinear Anomaly}},\ }\href {https://doi.org/10.1140/epjc/s10052-011-1665-7} {\bibfield  {journal} {\bibinfo  {journal} {Eur. Phys. J. C}\ }\textbf {\bibinfo {volume} {71}},\ \bibinfo {pages} {1665} (\bibinfo {year} {2011})},\ \Eprint {https://arxiv.org/abs/1007.4005} {arXiv:1007.4005 [hep-ph]} \BibitemShut {NoStop}%
\bibitem [{\citenamefont {Chiu}\ \emph {et~al.}(2012)\citenamefont {Chiu}, \citenamefont {Jain}, \citenamefont {Neill},\ and\ \citenamefont {Rothstein}}]{Chiu:2011qc}%
  \BibitemOpen
  \bibfield  {author} {\bibinfo {author} {\bibfnamefont {J.-y.}\ \bibnamefont {Chiu}}, \bibinfo {author} {\bibfnamefont {A.}~\bibnamefont {Jain}}, \bibinfo {author} {\bibfnamefont {D.}~\bibnamefont {Neill}},\ and\ \bibinfo {author} {\bibfnamefont {I.~Z.}\ \bibnamefont {Rothstein}},\ }\bibfield  {title} {\bibinfo {title} {{The Rapidity Renormalization Group}},\ }\href {https://doi.org/10.1103/PhysRevLett.108.151601} {\bibfield  {journal} {\bibinfo  {journal} {Phys. Rev. Lett.}\ }\textbf {\bibinfo {volume} {108}},\ \bibinfo {pages} {151601} (\bibinfo {year} {2012})},\ \Eprint {https://arxiv.org/abs/1104.0881} {arXiv:1104.0881 [hep-ph]} \BibitemShut {NoStop}%
\bibitem [{\citenamefont {Heinrich}\ \emph {et~al.}(2022)\citenamefont {Heinrich}, \citenamefont {Jahn}, \citenamefont {Jones}, \citenamefont {Kerner}, \citenamefont {Langer}, \citenamefont {Magerya}, \citenamefont {P\"oldaru}, \citenamefont {Schlenk},\ and\ \citenamefont {Villa}}]{Heinrich:2021dbf}%
  \BibitemOpen
  \bibfield  {author} {\bibinfo {author} {\bibfnamefont {G.}~\bibnamefont {Heinrich}}, \bibinfo {author} {\bibfnamefont {S.}~\bibnamefont {Jahn}}, \bibinfo {author} {\bibfnamefont {S.~P.}\ \bibnamefont {Jones}}, \bibinfo {author} {\bibfnamefont {M.}~\bibnamefont {Kerner}}, \bibinfo {author} {\bibfnamefont {F.}~\bibnamefont {Langer}}, \bibinfo {author} {\bibfnamefont {V.}~\bibnamefont {Magerya}}, \bibinfo {author} {\bibfnamefont {A.}~\bibnamefont {P\"oldaru}}, \bibinfo {author} {\bibfnamefont {J.}~\bibnamefont {Schlenk}},\ and\ \bibinfo {author} {\bibfnamefont {E.}~\bibnamefont {Villa}},\ }\bibfield  {title} {\bibinfo {title} {{Expansion by regions with pySecDec}},\ }\href {https://doi.org/10.1016/j.cpc.2021.108267} {\bibfield  {journal} {\bibinfo  {journal} {Comput. Phys. Commun.}\ }\textbf {\bibinfo {volume} {273}},\ \bibinfo {pages} {108267} (\bibinfo {year} {2022})},\ \Eprint {https://arxiv.org/abs/2108.10807} {arXiv:2108.10807 [hep-ph]} \BibitemShut {NoStop}%
\bibitem [{\citenamefont {Becher}\ \emph {et~al.}(2004)\citenamefont {Becher}, \citenamefont {Hill},\ and\ \citenamefont {Neubert}}]{Becher:2003qh}%
  \BibitemOpen
  \bibfield  {author} {\bibinfo {author} {\bibfnamefont {T.}~\bibnamefont {Becher}}, \bibinfo {author} {\bibfnamefont {R.~J.}\ \bibnamefont {Hill}},\ and\ \bibinfo {author} {\bibfnamefont {M.}~\bibnamefont {Neubert}},\ }\bibfield  {title} {\bibinfo {title} {{Soft collinear messengers: A New mode in soft collinear effective theory}},\ }\href {https://doi.org/10.1103/PhysRevD.69.054017} {\bibfield  {journal} {\bibinfo  {journal} {Phys. Rev. D}\ }\textbf {\bibinfo {volume} {69}},\ \bibinfo {pages} {054017} (\bibinfo {year} {2004})},\ \Eprint {https://arxiv.org/abs/hep-ph/0308122} {arXiv:hep-ph/0308122} \BibitemShut {NoStop}%
\bibitem [{\citenamefont {Bern}\ \emph {et~al.}(1994)\citenamefont {Bern}, \citenamefont {Dixon},\ and\ \citenamefont {Kosower}}]{Bern:1993kr}%
  \BibitemOpen
  \bibfield  {author} {\bibinfo {author} {\bibfnamefont {Z.}~\bibnamefont {Bern}}, \bibinfo {author} {\bibfnamefont {L.~J.}\ \bibnamefont {Dixon}},\ and\ \bibinfo {author} {\bibfnamefont {D.~A.}\ \bibnamefont {Kosower}},\ }\bibfield  {title} {\bibinfo {title} {{Dimensionally regulated pentagon integrals}},\ }\href {https://doi.org/10.1016/0550-3213(94)90398-0} {\bibfield  {journal} {\bibinfo  {journal} {Nucl. Phys. B}\ }\textbf {\bibinfo {volume} {412}},\ \bibinfo {pages} {751} (\bibinfo {year} {1994})},\ \Eprint {https://arxiv.org/abs/hep-ph/9306240} {arXiv:hep-ph/9306240} \BibitemShut {NoStop}%
\bibitem [{\citenamefont {Usyukina}\ and\ \citenamefont {Davydychev}(1994)}]{Usyukina:1994iw}%
  \BibitemOpen
  \bibfield  {author} {\bibinfo {author} {\bibfnamefont {N.~I.}\ \bibnamefont {Usyukina}}\ and\ \bibinfo {author} {\bibfnamefont {A.~I.}\ \bibnamefont {Davydychev}},\ }\bibfield  {title} {\bibinfo {title} {{New results for two loop off-shell three point diagrams}},\ }\href {https://doi.org/10.1016/0370-2693(94)90874-5} {\bibfield  {journal} {\bibinfo  {journal} {Phys. Lett. B}\ }\textbf {\bibinfo {volume} {332}},\ \bibinfo {pages} {159} (\bibinfo {year} {1994})},\ \Eprint {https://arxiv.org/abs/hep-ph/9402223} {arXiv:hep-ph/9402223} \BibitemShut {NoStop}%
\bibitem [{\citenamefont {Lee}\ and\ \citenamefont {Pomeransky}(2013)}]{Lee:2013hzt}%
  \BibitemOpen
  \bibfield  {author} {\bibinfo {author} {\bibfnamefont {R.~N.}\ \bibnamefont {Lee}}\ and\ \bibinfo {author} {\bibfnamefont {A.~A.}\ \bibnamefont {Pomeransky}},\ }\bibfield  {title} {\bibinfo {title} {{Critical points and number of master integrals}},\ }\href {https://doi.org/10.1007/JHEP11(2013)165} {\bibfield  {journal} {\bibinfo  {journal} {JHEP}\ }\textbf {\bibinfo {volume} {11}},\ \bibinfo {pages} {165}},\ \Eprint {https://arxiv.org/abs/1308.6676} {arXiv:1308.6676 [hep-ph]} \BibitemShut {NoStop}%
\bibitem [{\citenamefont {Pak}\ and\ \citenamefont {Smirnov}(2011)}]{Pak:2010pt}%
  \BibitemOpen
  \bibfield  {author} {\bibinfo {author} {\bibfnamefont {A.}~\bibnamefont {Pak}}\ and\ \bibinfo {author} {\bibfnamefont {A.}~\bibnamefont {Smirnov}},\ }\bibfield  {title} {\bibinfo {title} {{Geometric approach to asymptotic expansion of Feynman integrals}},\ }\href {https://doi.org/10.1140/epjc/s10052-011-1626-1} {\bibfield  {journal} {\bibinfo  {journal} {Eur. Phys. J. C}\ }\textbf {\bibinfo {volume} {71}},\ \bibinfo {pages} {1626} (\bibinfo {year} {2011})},\ \Eprint {https://arxiv.org/abs/1011.4863} {arXiv:1011.4863 [hep-ph]} \BibitemShut {NoStop}%
\bibitem [{\citenamefont {Becher}\ and\ \citenamefont {Bell}(2012)}]{Becher:2011dz}%
  \BibitemOpen
  \bibfield  {author} {\bibinfo {author} {\bibfnamefont {T.}~\bibnamefont {Becher}}\ and\ \bibinfo {author} {\bibfnamefont {G.}~\bibnamefont {Bell}},\ }\bibfield  {title} {\bibinfo {title} {{Analytic Regularization in Soft-Collinear Effective Theory}},\ }\href {https://doi.org/10.1016/j.physletb.2012.05.016} {\bibfield  {journal} {\bibinfo  {journal} {Phys. Lett. B}\ }\textbf {\bibinfo {volume} {713}},\ \bibinfo {pages} {41} (\bibinfo {year} {2012})},\ \Eprint {https://arxiv.org/abs/1112.3907} {arXiv:1112.3907 [hep-ph]} \BibitemShut {NoStop}%
\bibitem [{\citenamefont {Rothstein}\ and\ \citenamefont {Stewart}(2016)}]{Rothstein:2016bsq}%
  \BibitemOpen
  \bibfield  {author} {\bibinfo {author} {\bibfnamefont {I.~Z.}\ \bibnamefont {Rothstein}}\ and\ \bibinfo {author} {\bibfnamefont {I.~W.}\ \bibnamefont {Stewart}},\ }\bibfield  {title} {\bibinfo {title} {{An Effective Field Theory for Forward Scattering and Factorization Violation}},\ }\href {https://doi.org/10.1007/JHEP08(2016)025} {\bibfield  {journal} {\bibinfo  {journal} {JHEP}\ }\textbf {\bibinfo {volume} {08}},\ \bibinfo {pages} {025}},\ \Eprint {https://arxiv.org/abs/1601.04695} {arXiv:1601.04695 [hep-ph]} \BibitemShut {NoStop}%
\bibitem [{\citenamefont {Moult}\ \emph {et~al.}(2023)\citenamefont {Moult}, \citenamefont {Raman}, \citenamefont {Ridgway},\ and\ \citenamefont {Stewart}}]{Moult:2022lfy}%
  \BibitemOpen
  \bibfield  {author} {\bibinfo {author} {\bibfnamefont {I.}~\bibnamefont {Moult}}, \bibinfo {author} {\bibfnamefont {S.}~\bibnamefont {Raman}}, \bibinfo {author} {\bibfnamefont {G.}~\bibnamefont {Ridgway}},\ and\ \bibinfo {author} {\bibfnamefont {I.~W.}\ \bibnamefont {Stewart}},\ }\bibfield  {title} {\bibinfo {title} {{Anomalous dimensions from soft Regge constants}},\ }\href {https://doi.org/10.1007/JHEP05(2023)025} {\bibfield  {journal} {\bibinfo  {journal} {JHEP}\ }\textbf {\bibinfo {volume} {05}},\ \bibinfo {pages} {025}},\ \Eprint {https://arxiv.org/abs/2207.02859} {arXiv:2207.02859 [hep-ph]} \BibitemShut {NoStop}%
\bibitem [{\citenamefont {Banfi}\ \emph {et~al.}(2022)\citenamefont {Banfi}, \citenamefont {Dreyer},\ and\ \citenamefont {Monni}}]{Banfi:2021xzn}%
  \BibitemOpen
  \bibfield  {author} {\bibinfo {author} {\bibfnamefont {A.}~\bibnamefont {Banfi}}, \bibinfo {author} {\bibfnamefont {F.~A.}\ \bibnamefont {Dreyer}},\ and\ \bibinfo {author} {\bibfnamefont {P.~F.}\ \bibnamefont {Monni}},\ }\bibfield  {title} {\bibinfo {title} {{Higher-order non-global logarithms from jet calculus}},\ }\href {https://doi.org/10.1007/JHEP03(2022)135} {\bibfield  {journal} {\bibinfo  {journal} {JHEP}\ }\textbf {\bibinfo {volume} {03}},\ \bibinfo {pages} {135}},\ \Eprint {https://arxiv.org/abs/2111.02413} {arXiv:2111.02413 [hep-ph]} \BibitemShut {NoStop}%
\bibitem [{\citenamefont {Becher}\ \emph {et~al.}(2024)\citenamefont {Becher}, \citenamefont {Schalch},\ and\ \citenamefont {Xu}}]{Becher:2023vrh}%
  \BibitemOpen
  \bibfield  {author} {\bibinfo {author} {\bibfnamefont {T.}~\bibnamefont {Becher}}, \bibinfo {author} {\bibfnamefont {N.}~\bibnamefont {Schalch}},\ and\ \bibinfo {author} {\bibfnamefont {X.}~\bibnamefont {Xu}},\ }\bibfield  {title} {\bibinfo {title} {{Resummation of Next-to-Leading Nonglobal Logarithms at the LHC}},\ }\href {https://doi.org/10.1103/PhysRevLett.132.081602} {\bibfield  {journal} {\bibinfo  {journal} {Phys. Rev. Lett.}\ }\textbf {\bibinfo {volume} {132}},\ \bibinfo {pages} {081602} (\bibinfo {year} {2024})},\ \Eprint {https://arxiv.org/abs/2307.02283} {arXiv:2307.02283 [hep-ph]} \BibitemShut {NoStop}%
\bibitem [{\citenamefont {Ferrario~Ravasio}\ \emph {et~al.}(2023)\citenamefont {Ferrario~Ravasio}, \citenamefont {Hamilton}, \citenamefont {Karlberg}, \citenamefont {Salam}, \citenamefont {Scyboz},\ and\ \citenamefont {Soyez}}]{FerrarioRavasio:2023kyg}%
  \BibitemOpen
  \bibfield  {author} {\bibinfo {author} {\bibfnamefont {S.}~\bibnamefont {Ferrario~Ravasio}}, \bibinfo {author} {\bibfnamefont {K.}~\bibnamefont {Hamilton}}, \bibinfo {author} {\bibfnamefont {A.}~\bibnamefont {Karlberg}}, \bibinfo {author} {\bibfnamefont {G.~P.}\ \bibnamefont {Salam}}, \bibinfo {author} {\bibfnamefont {L.}~\bibnamefont {Scyboz}},\ and\ \bibinfo {author} {\bibfnamefont {G.}~\bibnamefont {Soyez}},\ }\bibfield  {title} {\bibinfo {title} {{Parton Showering with Higher Logarithmic Accuracy for Soft Emissions}},\ }\href {https://doi.org/10.1103/PhysRevLett.131.161906} {\bibfield  {journal} {\bibinfo  {journal} {Phys. Rev. Lett.}\ }\textbf {\bibinfo {volume} {131}},\ \bibinfo {pages} {161906} (\bibinfo {year} {2023})},\ \Eprint {https://arxiv.org/abs/2307.11142} {arXiv:2307.11142 [hep-ph]} \BibitemShut {NoStop}%
\bibitem [{\citenamefont {Nagy}\ and\ \citenamefont {Soper}(2007)}]{Nagy:2007ty}%
  \BibitemOpen
  \bibfield  {author} {\bibinfo {author} {\bibfnamefont {Z.}~\bibnamefont {Nagy}}\ and\ \bibinfo {author} {\bibfnamefont {D.~E.}\ \bibnamefont {Soper}},\ }\bibfield  {title} {\bibinfo {title} {{Parton showers with quantum interference}},\ }\href {https://doi.org/10.1088/1126-6708/2007/09/114} {\bibfield  {journal} {\bibinfo  {journal} {JHEP}\ }\textbf {\bibinfo {volume} {09}},\ \bibinfo {pages} {114}},\ \Eprint {https://arxiv.org/abs/0706.0017} {arXiv:0706.0017 [hep-ph]} \BibitemShut {NoStop}%
\bibitem [{\citenamefont {Nagy}\ and\ \citenamefont {Soper}(2019)}]{Nagy:2019pjp}%
  \BibitemOpen
  \bibfield  {author} {\bibinfo {author} {\bibfnamefont {Z.}~\bibnamefont {Nagy}}\ and\ \bibinfo {author} {\bibfnamefont {D.~E.}\ \bibnamefont {Soper}},\ }\bibfield  {title} {\bibinfo {title} {{Parton showers with more exact color evolution}},\ }\href {https://doi.org/10.1103/PhysRevD.99.054009} {\bibfield  {journal} {\bibinfo  {journal} {Phys. Rev. D}\ }\textbf {\bibinfo {volume} {99}},\ \bibinfo {pages} {054009} (\bibinfo {year} {2019})},\ \Eprint {https://arxiv.org/abs/1902.02105} {arXiv:1902.02105 [hep-ph]} \BibitemShut {NoStop}%
\bibitem [{\citenamefont {Forshaw}\ \emph {et~al.}(2019)\citenamefont {Forshaw}, \citenamefont {Holguin},\ and\ \citenamefont {Pl\"atzer}}]{Forshaw:2019ver}%
  \BibitemOpen
  \bibfield  {author} {\bibinfo {author} {\bibfnamefont {J.~R.}\ \bibnamefont {Forshaw}}, \bibinfo {author} {\bibfnamefont {J.}~\bibnamefont {Holguin}},\ and\ \bibinfo {author} {\bibfnamefont {S.}~\bibnamefont {Pl\"atzer}},\ }\bibfield  {title} {\bibinfo {title} {{Parton branching at amplitude level}},\ }\href {https://doi.org/10.1007/JHEP08(2019)145} {\bibfield  {journal} {\bibinfo  {journal} {JHEP}\ }\textbf {\bibinfo {volume} {08}},\ \bibinfo {pages} {145}},\ \Eprint {https://arxiv.org/abs/1905.08686} {arXiv:1905.08686 [hep-ph]} \BibitemShut {NoStop}%
\bibitem [{\citenamefont {De~Angelis}\ \emph {et~al.}(2021)\citenamefont {De~Angelis}, \citenamefont {Forshaw},\ and\ \citenamefont {Pl\"atzer}}]{DeAngelis:2020rvq}%
  \BibitemOpen
  \bibfield  {author} {\bibinfo {author} {\bibfnamefont {M.}~\bibnamefont {De~Angelis}}, \bibinfo {author} {\bibfnamefont {J.~R.}\ \bibnamefont {Forshaw}},\ and\ \bibinfo {author} {\bibfnamefont {S.}~\bibnamefont {Pl\"atzer}},\ }\bibfield  {title} {\bibinfo {title} {{Resummation and Simulation of Soft Gluon Effects beyond Leading Color}},\ }\href {https://doi.org/10.1103/PhysRevLett.126.112001} {\bibfield  {journal} {\bibinfo  {journal} {Phys. Rev. Lett.}\ }\textbf {\bibinfo {volume} {126}},\ \bibinfo {pages} {112001} (\bibinfo {year} {2021})},\ \Eprint {https://arxiv.org/abs/2007.09648} {arXiv:2007.09648 [hep-ph]} \BibitemShut {NoStop}%
\end{thebibliography}%

\end{document}